\documentclass[11pt]{article}
\usepackage{moriond,epsfig}

\def\be{\begin{equation}}
\def\ee{\end{equation}}
\def\bea{\begin{eqnarray}}
\def\eea{\end{eqnarray}}
\def\etal{{\it et al.\ }}
\def\kmps{${\rm\thinspace
km}{\rm\thinspace s}^{-1}$} 

\begin{document}
\vspace*{4cm}
\title{THE DIPOLE ANISOTROPY OF THE 2MASS REDSHIFT SURVEY}

\author{P\.IR\.IN ERDO\u{G}DU AND THE 2MRS TEAM\footnote{J.P. Huchra (CfA),
O. Lahav (UCL), M. Colless (AAO), R.M. Cutri (IPAC), E. Falco
(Harvard-Smithsonian), T. George (CFTC), T. Jarrett (IPAC), D. H. Jones (AAO),
C.S. Kochanek (Ohio State), L. Macri (NOAO), J. Mader (Keck), N. Martimbeau
(CfA), M. Pahre (CfA), Q. Parker (Maquarie), A. Rassat (UCL), W. Saunders (AAO, Maquarie)}}
\address{School of Physics \& Astronomy, University of Nottingham,
University Park, Nottingham, NG7 2RD, UK}

\maketitle\abstracts{We estimate the f\mbox{}l\mbox{}ux weighted  
acceleration on the Local Group (LG) from the near-infrared Two
Micron All Sky Redshift Survey (2MRS).
The near-infrared f\mbox{}l\mbox{}ux weighted dipoles are very robust because they closely approximate a mass weighted dipole, bypassing 
the effects of redshift distortions and require no preferred reference frame. 
We use this method with the redshift information to determine the
change in dipole with distance.  
The LG dipole seemingly converges by 60 $h^{-1} {\rm Mpc}$. 
Assuming convergence, the comparison of the 2MRS f\mbox{}l\mbox{}ux dipole and the CMB dipole 
provides a value for  
the combination of the mass density  
and luminosity bias parameters  $\Omega_{\rm m}^{0.6}/b_{\rm L}= 0.40 \pm 0.09$}.

\section{Introduction}
The most popular mechanism for the formation of large-scale structure
and motions in the Universe is the gravitational growth of primordial
density perturbations.  According to this paradigm, the peculiar acceleration vector ${\bf g}({\bf r})$ is induced by
the matter distribution around position ${\bf r}$ and if the density
perturbations are small enough to be approximated by a linear theory,
then the peculiar velocity
field, ${\bf v}({\bf r})$, is proportional to the peculiar acceleration:
\be
{\bf v}({\bf r})= \frac{H_0 f(\Omega_{\rm m})}{4 \pi G \bar{\rho}}
{\bf g}({\bf r})=\frac{2f(\Omega_{\rm m})}{3 H_0\Omega_{\rm m}}{\bf
g}({\bf r}),
\label{eqn:v(r)}
\ee
where $H_0$ = 100 $h$ ${\rm km} {\rm s}^{-1} {\rm Mpc}^{-1}$ is the
Hubble constant and $f(\Omega_{\rm m})\simeq\Omega_{\rm m}^{0.6}$ is
the logarithmic derivative of the amplitude of the growing mode of the
perturbations in mass with respect to the scale factor (Peebles 1980).

It is now widely accepted
that the dipole anisotropy of the cosmic microwave background (CMB) is
a direct and accurate measurement of the LG peculiar velocity.
The LG acceleration can also be
estimated using surveys of the galaxies tracing the density
inhomogeneities responsible for the acceleration. If the mass can be related to light by a bias parameter, $b_L$, then by comparing the CMB
velocity vector with the acceleration vector 
obtained from the
galaxy surveys, it is possible to investigate the cause of the LG
motion and its cosmological implications.

Like peculiar acceleration, the flux of light received from a galaxy falls off inversely as the square of the distance. If the mass-to-light ratio is constant, we can relate the two by
\bea
{\bf g}({\bf r})&=& {\rm G} \sum\limits_{i}\,
M_{i} \frac{\hat{{\bf r}}_i}{r_{i}^2} \simeq {\rm G}
\left\langle\frac{M}{L}\right\rangle\sum\limits_{i}\, L_{i}
\frac{\hat{{\bf r}}_i}{r_{i}^2}\nonumber \\ &=&4\pi{\rm G}
\left\langle\frac{M}{L}\right\rangle\sum\limits_{i}\, {\rm S}_{i}
\hat{{\bf r}}_i,
\label{eqn:glum}
\eea
where the sum is over all galaxies in the Universe,
$\left\langle M/L \right\rangle$ is the average mass-to-light ratio
and ${\rm S}_i=L_i/4 \pi r^2$ is the f\mbox{}l\mbox{}ux of galaxy $i$.
The peculiar
velocity vector is derived by substituting Equation~\ref{eqn:glum} into the
second line of Equation~\ref{eqn:v(r)}.
For a f\mbox{}l\mbox{}ux limited catalogue
the observed LG velocity is
\be
{\bf v}({\bf r})= \frac{8\pi{\rm G}f(\Omega_{\rm m})}{3 H_0\Omega_{\rm
m}b_{\rm L}} \left\langle\frac{M}{L}\right\rangle\sum\limits_{i}\,
w_{L_i}{\rm S}_{i} \hat{{\bf r}}_i
\label{eqn:vl}
\ee
where $b_{\rm L}$ is the luminosity bias factor introduced to account for the
dark matter haloes not fully represented by 2MRS galaxies and $0\leq w_{L_i}
\leq 1$
is the weight assigned to galaxy $i$ to account for the 
 luminosity that was not observed due to the flux limit of the survey.
With the inclusion
of redshift information, we can calculate Equation~\ref{eqn:vl}
within concentric spheres with increasing radii 
and thus estimate the distance at which most of the LG velocity is generated
generated ({\it the convergence depth}). 

\section{The Two Micron All-Sky Redshift Survey}\label{sec:dip:data}
\begin{figure*}
\psfig{figure=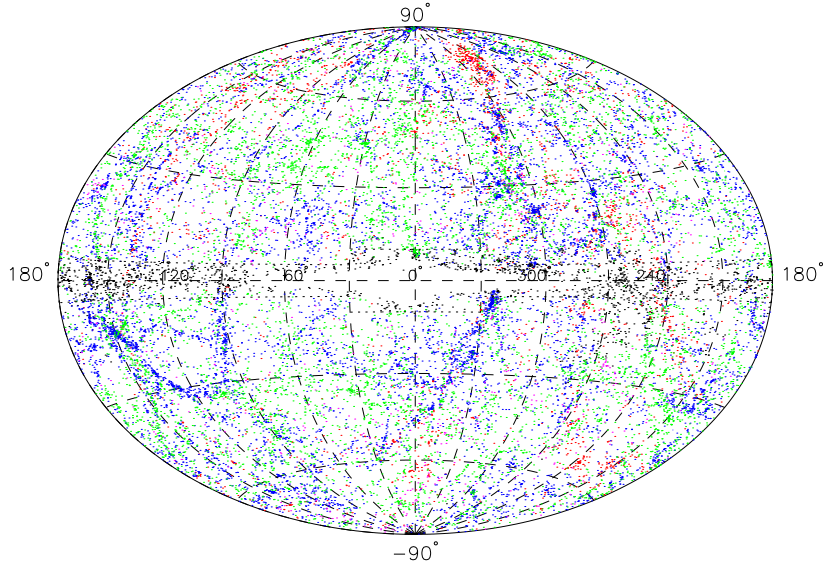, height=9cm}
\caption{All Objects in the 2MASS Redshift Catalogue in
Galactic Aitoff Projection. Galaxies with ${\rm z}\leq0.01$ are
plotted in red, $0.01<{\rm z}\le0.025$ are plotted in blue,
$0.025<{\rm z}<0.05$ are plotted in green and ${\rm z}\geq0.05$ are
plotted in magenta. Galaxies without measured redshifts are plotted in
black. The masked region is outlined by dashed lines.
\label{fig:2MRS}}
\end{figure*}

The Two Micron All-Sky Redshift Survey (Huchra \etal 2005, Erdo\u{g}du \etal 2006) 
is the densest all-sky
redshift survey to date. The first phase of 2MRS is now
complete. In this phase we obtained redshifts for approximately
23,150 2MASS galaxies from a total sample of 24,773 galaxies with
extinction corrected magnitudes (Schlegel, Finkbeiner \& Davis 1998)
brighter than $K_{\rm s}=11.25$.
This magnitude limit corresponds to a
median redshift of $z\simeq0.02$ ($\simeq$ 6000 \kmps).  
Figure~\ref{fig:2MRS} shows all the objects in 2MRS in
Galactic Aitoff Projection. Galaxies with ${\rm z}\leq0.01$ are
plotted in red, $0.01<{\rm z}\le0.025$ are plotted in blue,
$0.025<{\rm z}<0.05$ are plotted in green and ${\rm z}\geq0.05$ are
plotted in magenta. Galaxies without measured redshifts (around 1600) 
are plotted in
black. 

The 2MRS sample
has very good photometric uniformity and an unprecedented
integral sky coverage. The photometric uniformity is better than
$4\%$ over the sky including the celestial poles
and the survey is essentially complete
 down to very low galactic latitudes (Huchra \etal 2005). 
In order to account for incompleteness at these galactic latitudes
we fill the plane with galaxies sampled from adjacent longitude/distance bins.

\section{Results}

\begin{figure*}
\psfig{figure=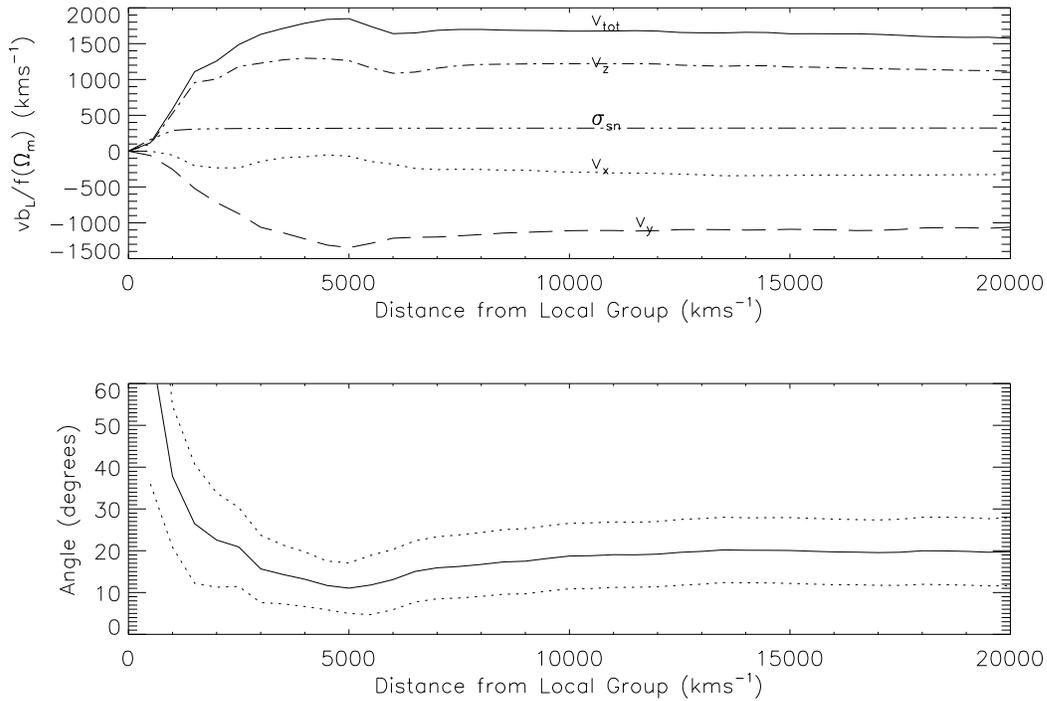,height=10cm}
\caption{{\bf Top}: Three components and the magnitudes of the acceleration of the
Local Group due to galaxies within a series of successively larger
concentric spheres centred on the local group. 
{\bf Bottom}: Convergence of the direction of the LG dipole where the misalignment
angle is between the LG and the CMB dipoles. The dotted lines denote
1$\sigma$ errors from shot noise.
\label{fig:dipole}} 
\end{figure*}

Figure~\ref{fig:dipole} shows the three components and the magnitudes of the acceleration of the
Local Group due to galaxies within a series of successively larger
concentric spheres centred on the local group (top plot). The bottom plot shows  convergence of the direction of the LG dipole where the misalignment
angle is between the LG and the CMB dipoles ($v_{LG}=627\pm22$
\kmps, towards $l_{LG}=273^\circ\pm3^\circ$,
$b_{LG}=29^\circ\pm3^\circ$, Bennett \etal 2003, Courteau \& Van Den Bergh, 1999). 

It is evident in the top plot that 
the LG velocity is dominated by structures within a
distance of 6000 \kmps.
The `tug of war' between the Great
Attractor and the Perseus-Pisces is clearly evident. The dip in the
velocity vector is an indication that the local flow towards the Great
Attractor\footnote{ By `Great Attractor', it is meant the entire
steradian on the sky centred at ($l\sim310^\circ$,$b\sim20^\circ$)
covering a distance of 20 $h^{-1}$ Mpc to 60 $h^{-1}$ Mpc.} is
counteracted by the Perseus-Pisces complex in the opposite
direction. The dipole seems to converge by 6000 \kmps.

The misalignment angle between the LG and the CMB dipole is smallest at
5000 \kmps where it drops to 12$^\circ\pm$7$^\circ$ 
and increases slightly at larger distances presumably due to shot-noise. 
The direction of the flux dipole (l=251$^\circ\pm$12$^\circ$,b=37$^\circ\pm$ 10  $^\circ$) is in good agreement with the 2MASS dipole derived by Maller \etal 
(2003).
The difference in results is probably due to the fact that they use 
a higher latitude cutoff in the mask ($|b|<7^\circ$) 
and exclude all galaxies below this latitude. We confirm this by changing our 
treatment of the Zone of Avoidance to match theirs. We find that 
the flux dipole is very close to their dipole direction. 
Their
limiting Kron magnitude is $K_s=13.57$ which corresponds to an
effective depth of 200 $h^{-1}$ Mpc. As their sample is deep enough to pick
out galaxies in the Shapley Supercluster, the comparison of their dipole
value with our values suggests that the contribution to 
the LG dipole from structure further away 
than the maximum distance of our analysis is not significant. 

Assuming convergence, we equate the velocity
inferred from the CMB measurements with the value derived from a
galaxy survey and obtain a value for
the combination of the mass density  
and luminosity bias parameters:  
$\Omega_{\rm m}^{0.6}/b_{\rm L}= 0.40 \pm 0.09$.
If we adopt $\Omega_{\rm m}=0.23$ from the WMAP3 results (Spergel \etal 2006) we get 
$b_L\simeq1$. This suggest that the 2MRS galaxies are unbiased.

A detailed description of the analysis and results 
outlined in this paper can be found in Erdo\u{g}du \etal (2006).

\section*{Acknowledgments}
OL acknowledges a PPARC Senior
Research Fellowship.  JPH, LM, CSK, NM, and TJ are supported by NSF
grant AST-0406906, and EF's research is partially supported by the Smithsonian Institution.  
DHJ is supported as a Research Associate by
Australian Research Council Discovery-Projects Grant (DP-0208876),
administered by the Australian National University.  This publication
makes use of data products from the Two Micron All Sky Survey, which
is a joint project of the University of Massachusetts and the Infrared
Processing and Analysis Center/California Institute of Technology,
funded by the National Aeronautics and Space Administration and the
National Science Foundation. This research has also made use of the NASA/IPAC Extragalactic Database (NED) 
which is operated by the Jet Propulsion Laboratory, California Institute of Technology, under contract 
with the National Aeronautics and Space Administration and the SIMBAD database,
operated at CDS, Strasbourg, France.

\end{document}